\documentclass[onecolumn,aps,nofootinbib,superscriptaddress,12pt]{revtex4-1}
\usepackage{amssymb,mathtools, graphicx, enumerate, fullpage, subfig}
\usepackage[super]{nth}
\usepackage{color}
\usepackage{titlesec}

\newcommand{\eq}[1]{Eq.\,\eqref{eq:#1}}
\newcommand{\fig}[1]{Fig.\,\ref{fig:#1}}

\begin{document}
\title{Reconciling vacuum laser acceleration theory and experiment}
\author{B. Manuel Hegelich}
\affiliation{Center for High Energy Density Science, University of Texas, Austin, Texas, 78712}
\author{Lance A. Labun}
\affiliation{Center for High Energy Density Science, University of Texas, Austin, Texas, 78712}
\author{Ou Z. Labun}
\email{Authors alphabetical}
\affiliation{Center for High Energy Density Science, University of Texas, Austin, Texas, 78712}
\begin{abstract}
The classical theory of single-electron dynamics in focused laser pulses is the foundation of both the relativistic ponderomotive force (RPF), which in turn underlies models of laser-collective-plasma dynamics, and the discovery of novel strong-field radiation dynamics.  Despite this bedrock importance, consensus eludes the community as to whether acceleration of single electrons in vacuum has been observed in experiment.  We analyze the experiment of Malka et al. (1998) with respect to several features that were neglected in modeling and that can restore consistency between theory predictions and experimental data.  The right or wrong pulse profile function, laser parameters, or initial electron distribution each can make or break the agreement between predictions and data.  The laser phase at which the electron's interaction with the pulse begins has a large effect, explaining why much larger energies are achieved by electrons liberated in the focal region by photoionization from high-Z atoms and by electrons ejected from a plasma mirror.  Finally we estimate the error in a typical electron spectrum arising from fluctuating focal spot size in state-of-the-art ultra-relativistic laser facilities.  Our results emphasize the importance of thoroughly characterizing laser parameters in order to achieve quantitatively accurate predictions and the precision required for discovery science.
\end{abstract}
\maketitle

\section{Introduction}

Experiments have tested our understanding single-electron acceleration by lasers in vacuum up to (peak) intensities of $I\simeq 10^{20}$ W/cm$^2$ \cite{Malka:1997zwk,mcnaught1998photoelectron, moore1999laser, dichiara2008relativistic,payeur2012generation, cline2013first,carbajo2015direct, thevenet2016vacuum}.  An early experiment \cite{Malka:1997zwk} provoked disagreement as to whether the data was consistent with the theory of single-electron motion in a laser \cite{McDonald:1998pu,Quesnel:1998zz}.   Later experiments have either not been conclusive \cite{cline2013first,carbajo2015direct} or not tested comparable conditions \cite{mcnaught1998photoelectron, moore1999laser, dichiara2008relativistic,payeur2012generation, thevenet2016vacuum}.  To our knowledge, no work has addressed how the experimental data can be consistent with the theoretical understanding of particle dynamics in a focused laser.  

Determining the causes of the theory-experiment disagreement remains important, despite the time elapsed, in order to reduce future battles over whether or not an experiment has detected a new effect.  The physics of single-electron-laser scattering is essential to many discovery goals, from theories of strong-field ionization \cite{mcnaught1998photoelectron, moore1999laser, dichiara2008relativistic,payeur2012generation} to strong-field corrections to particle dynamics from quantum electrodynamics \cite{Bula:1996st,Bamber:1999zt} and classical radiation reaction \cite{cole2018experimental,poder2018experimental}.

Additionally, the relativistic ponderomotive force (RPF) is thought to be valid around $I\simeq 10^{20}$ W/cm$^2$ \cite{Quesnel:1998zz}, and therefore relevant to explaining electron acceleration in vacuum.  The RPF now underlies models of laser acceleration in more general plasma conditions \cite{beg1997study,key1998hot,wilks2001energetic,haines2009hot,yang2011explicit}.  Broad use of the RPF begs the question why it seems to fail to describe the relatively simple experiment of Ref. \cite{Malka:1997zwk}.

We find that the experimental data of Ref. \cite{Malka:1997zwk} is consistent with theory when several previously omitted features are incorporated in the model of the experiment.  One feature, the possibility of a sudden turn-on of the laser-electron interaction, violates assumptions of the RPF and the theory analysis by Ref. \cite{McDonald:1998pu}.  The other three features are consistent with RPF theory and highlight subtleties in understanding and applying the RPF or the need for more rigorous experimental methods.  
We do not claim these four exhaust the physics effects that can render the theory and experiment consistent, but rather focus on lessons for the analysis of future experiments.  Since understanding the physics at work in an experiment implies plausibly excluding alternative explanations of the data, it suffices for us to exhibit a single counter-example--a single alternative explanation that reproduces the data--to prove that a given variable changes outcomes and must be measured or controlled better.

Beyond resolving an old disagreement, one of our goals here is to point to what laser parameters and initial conditions are important to measure accurately.  A significant difficulty for newer laser-plasma experiments is that parameters describing the laser field become increasingly difficult to measure at ultra-high intensity.  Pulse profiles are usually measured using a subaperture beam, often before full amplification and never at full pulse energy.  Focal spot sizes are measured without a target present and at reduced intensity, and laser energies are also inferred from percent-level fractions of the beam.  This reduction in knowledge increases uncertainty in predictions at higher intensity.  We will show that the same uncertainty in a parameter that hobbles analysis of Ref. \cite{Malka:1997zwk} translates to a much larger discrepancy in an observable as intensities rise to $I\sim 10^{23}$\,W/cm$^2$ and beyond.

\section{The disagreement}
The interpretation of the data in Ref. \cite{Malka:1997zwk} was criticized as being incompatible with current understanding of laser acceleration of electrons in vacuum.
In the experiment, an $I\simeq 10^{19}$ W/cm$^2$ laser interacted with low energy ($E_{\rm kin}=m_e(\gamma-1)\simeq 10$ keV) electrons and accelerated electrons to 200-900 keV, observed in the polarization plane but not perpendicular to it.  The angle and energy distributions were considered consistent with a simple model of the laser fields.  However, others argued that in the laser's parameter regime, electrons should not be able to gain energy \cite{McDonald:1998pu}, and an equal number of electrons should be observed in the plane perpendicular to laser polarization, because the RPF scatters electrons into an azimuthally symmetric distribution about the laser axis \cite{Quesnel:1998zz}.  

More recent experiments have also set out to measure energy gain by free electrons interacting with a laser in vacuum, but do not provide insight into the older experiment.  Both Refs. \cite{cline2013first,carbajo2015direct} report energy gain much smaller ($\Delta E/E\lesssim 0.05$) than reported by \cite{Malka:1997zwk} ($\Delta E/E\simeq 90$), and in fact the data provided in those works is insufficient to determine whether significant energy gain was detected.\footnote{In both Refs. \cite{cline2013first,carbajo2015direct} electrons initially have finite kinetic energy and significant dispersion in energy.  One qualitatively observes increased dispersion, showing that some electrons gain energy, but does not show that the average energy of the electron increases\cite{cline2013first}.  The other omits the data necessary to distinguish increased dispersion from an increased average energy \cite{carbajo2015direct}.}   Acceleration of photoionized electrons is not comparable, because the electron becomes free to move under influence of the laser field only at a finite time, typically close the peak laser intensity.  As we discuss below, this semi-infinite interaction time has an important impact on the electron dynamics and possibility of energy gain.  However, in one experiment, measurements in the plane perpendicular to polarization suggest that photoionized electrons may have been accelerated into a more azimuthally symmetric distribution \cite{kalashnikov2015diagnostics}.

Theory and experiment work has suggested that improved modeling of the electrons' initial momentum distribution could help explain the discrepancy \cite{chowdhury2004electron}, but dedicated modeling of the experiments' conditions is necessary to see whether or not a particular missing effect suffices to explain the data in Ref. \cite{Malka:1997zwk}.  Indeed, a significant reason for the disagreement is that theory contributions often work on generalities, simulate special cases, and therefore miss the impact of experiments' specific conditions.  

\section{Some missing pieces}

We investigate four features of the experiment that can help resolve the disagreement with theory:  (1) the laser pulse profile, (2) the time the interaction begins, (3) the angle (relative to the beam axis) at which the electrons are traveling before the interaction begins, and (4) shot-to-shot variation in the summary parameters of the laser pulse (energy, spot size, pulse duration).  These only sample the possible changes or correcitons to the model of the experiment; however each addresses a subtlety in transferring the theory to the interpretation of the data.  Each feature violates an assumption or changes an outcome from the reasoning presented in Refs. \cite{McDonald:1998pu} and \cite{Quesnel:1998zz}.  

\subsection{Laser pulse profile}

Ref. \cite{Malka:1997zwk} did not provide the temporal profile of the laser pulse in the experiment.  The authors modeled their pulse with a sine-squared function.  According to Ref. \cite{McDonald:1998pu}, the acceleration found in sine-squared and gaussian profiles is unphysical, arising from the fact that these profiles do not satisfy 
\begin{align}\label{eq:profilecondition}
\frac{d(\ln f)}{d\varphi}\ll 1,
\end{align}
a condition derived together with the gaussian beam spatial field distribution. This requires an at-most exponential dependence on $\varphi$; hence Ref. \cite{McDonald:1998pu} recommends hyperbolic secant.  In fact, consistency in the approximation scheme can be restored by including order $f'/f$ corrections in the gaussian beam fields \cite{Quesnel:1998zz}.  In order to compare to previous simulation results and discuss the impact of these modeling choices we simulate with three profiles:
\begin{align}\label{eq:profiles}
f_{I}(\varphi) &= \sin^2(\varphi/\varphi_0)\\
f_{II}(\varphi) &= \exp(-\varphi^2/\varphi_0^2)\\
f_{III}(\varphi) &= \mathrm{sech}(\varphi/\varphi_0),
\end{align}
where $\varphi=k_0(x-ct)$ is the dimensionless phase and $\varphi_0$ is linearly related to pulse duration, usually measured as the full-width half-max in intensity.  

\begin{figure}[h]
\includegraphics[width=0.6\textwidth]{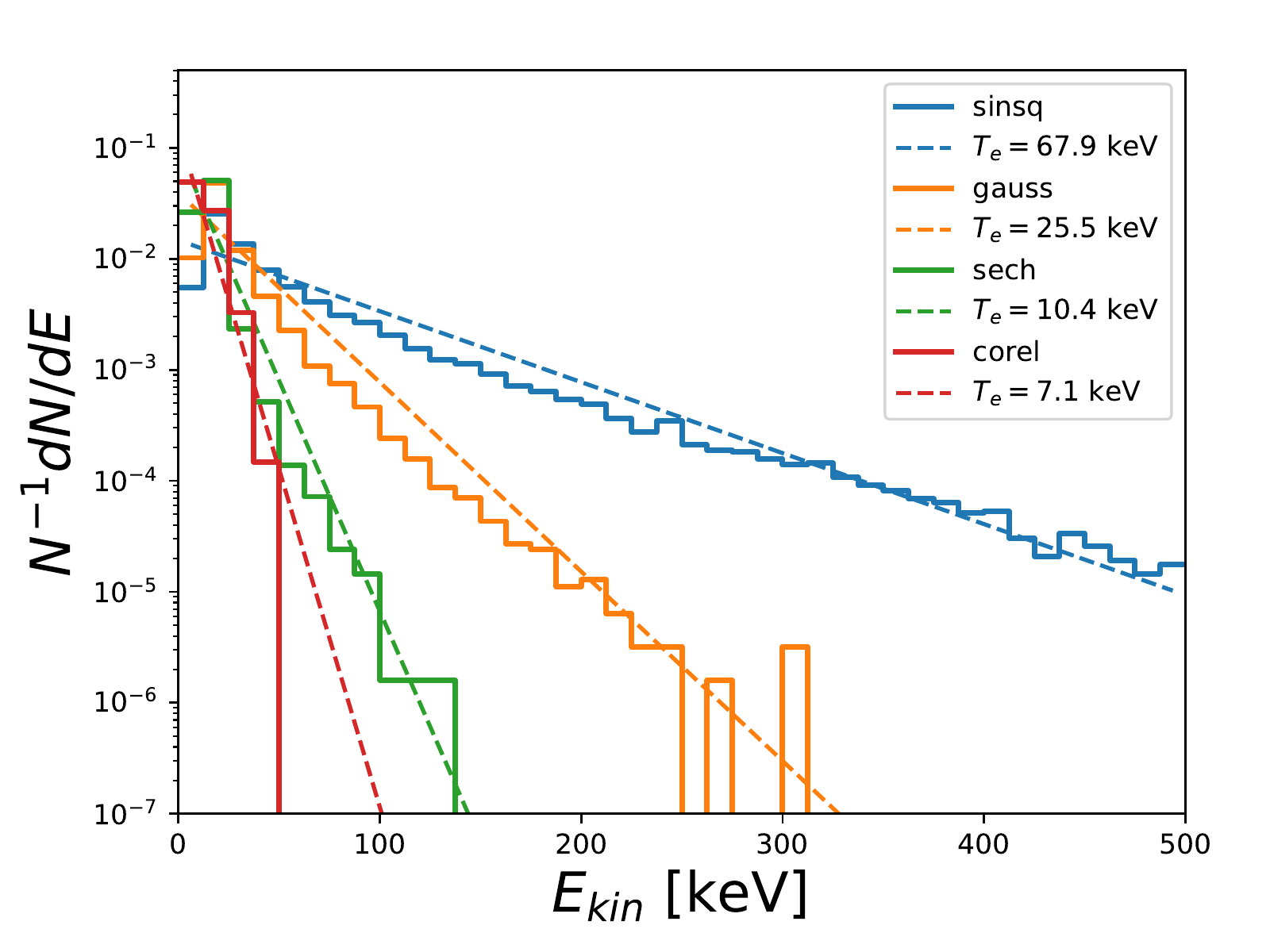}
\caption{\label{fig:profiles} Normalized electron energy distributions $N^{-1}(dN/dE)$ for sine-squared, gaussian, hyperbolic secant and three-peaked pulse measured at CoReLS.  Laser parameters are $a_0=3.0$, $w_0=10\,\mu$m and $\tau=350$ fs.  Maxwellian fits are plotted with the fit temperature shown in the key.}
\end{figure}

Fig. \ref{fig:profiles} compares the final energy distribution $dN/dE$ of electrons accelerated by four different temporal profiles, including sine-squared and the recommended hyperbolic secant.  The laser has peak intensity corresponding to $a_0=3$, focal spot radius $w_0=10$\,$\mu$,m, and full-width half-max duration (measured on intensity) of $\Delta\tau=350$ fs. The electrons are initially zero temperature with initial velocity parallel to the laser propagation axis, $\vec v=(0,0,0.2)c$, corresponding to kinetic energy $E_{\rm kin}=m(\gamma-1)=10.5$\,keV.  The electrons are initially uniformly distributed in a cubic 3-dimensional volume around the focal spot within $3w_0$ of the beam axis.\footnote{Increasing the volume covered by electron initial positions would result in more electrons receiving small or negligible acceleration and thereby increase the weight of $dN/dE$ around the initial kinetic energy without significantly affecting the fit at higher energies.}

Recalling that the mean energy per particle in a Maxwellian distribution is $\langle E\rangle/N = T$, we see that sine-squared and gaussian temporal profiles result in a significant gain in average energy.  Compared to the sech profile, gaussian and sine-squared profiles generally over-estimate energy gain \cite{McDonald:1998pu}.  While the hyperbolic secant does not significantly change the average energy, the electron distribution is significantly heated, evolving from a zero-temperature distribution to a finite temperature distribution.  Some electrons gain energy, while other electrons lose energy, showing that the reasoning of Ref. \cite{McDonald:1998pu} is incomplete.  This observation of heating without significant change in mean energy is consistent (at the provided, qualitative level) with a more recent experiment at $a_0\simeq 5$ and much longer pulse duration ($\Delta\tau\simeq 5$ ps) \cite{cline2013first}.

\begin{figure}[h]
\includegraphics[width=0.6\textwidth]{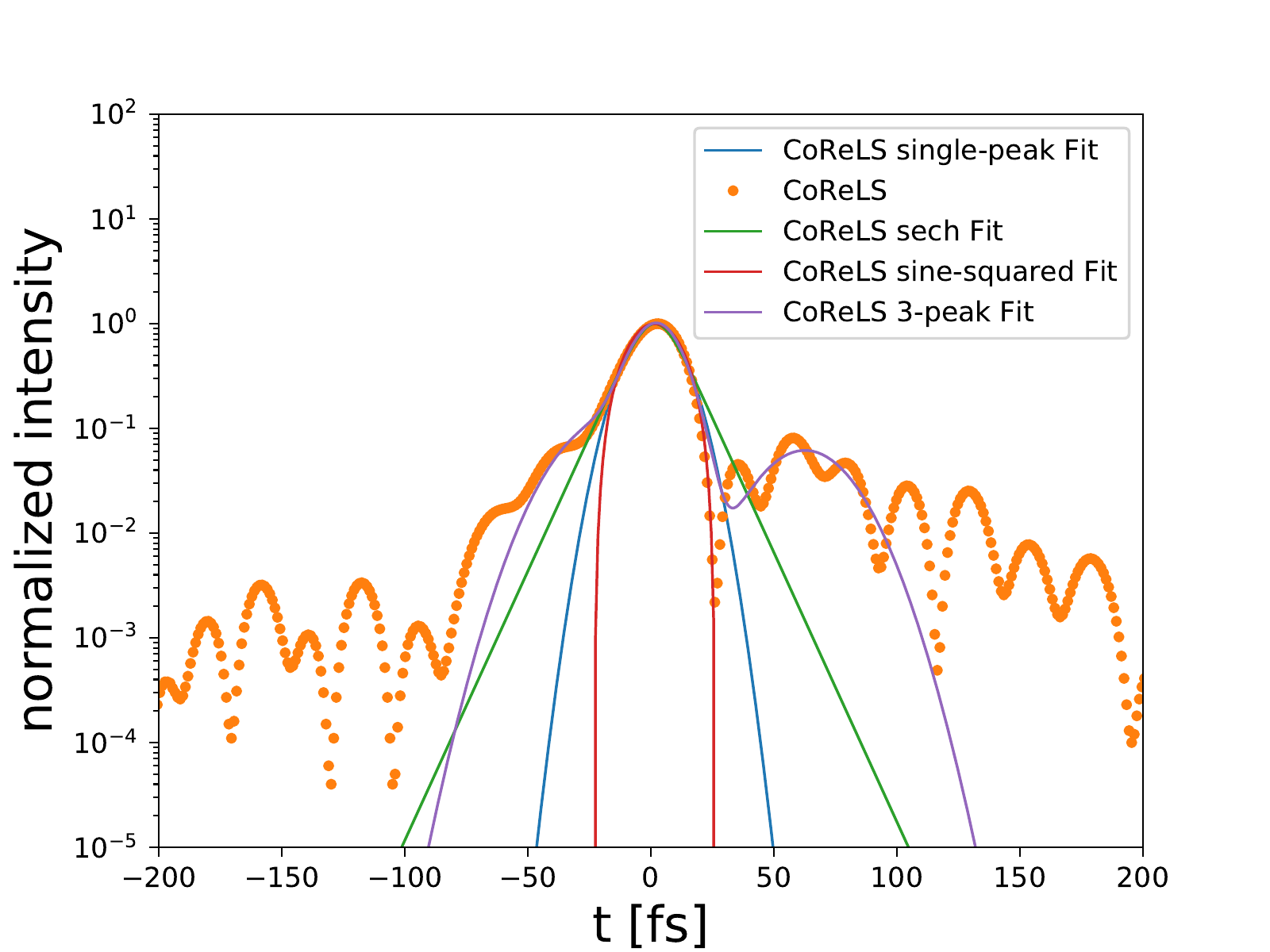}
\caption{\label{fig:profilefits} Temporal profile measured on CoReLS (points) compared to four possile fit functions.}
\end{figure}

Gaussian temporal profiles remain useful because measured laser pulse profiles are often fit passably by gaussians and not often fit by hyperbolic secant functions.  This fact is emphasized by the observation, on several ultra-high-intensity systems, of multiple peaks in the temporal profile.  In \fig{profilefits}, we fit a standard (single-peak) gaussian, a three-peak gaussian, a hyperbolic secant function and sine-squared function to the temporal profile of the 4PW Ti:Sapphire laser system at the Center for Relativistic Laser Science (CoReLS) in Gwangju, Korea.  The central peak is best fit by a gaussian, but a model with more degrees of freedom is clearly necessary to incorporate relevant features at $\sim 10\%$ peak intensity.  The CoReLS-labeled result in Fig. \ref{fig:profiles} used a 3-peaked gaussian model of a similar (not identical) measurement.  The three-peaked gaussian predicts even less energy gain and less heating than the hyperbolic secant.  Thus, measured temporal profiles may be closer to gaussian and contain other features essential to accurately predicting final electron energies.

Measuring the laser pulse profile obtained for each experiment is thus essential for analysis and interpretation of the data.  The temporal profile beyond the central  gaussian peak is important, making a second-order autocorrelation measurement insufficient.  Many systems, especially Ti:Sa systems, do not perform on-shot measurements of every system shot but rather use scanning measurements during the laser alignment procedure,  i.e. averaging over 10,000 or more pulses.  A temporal profile derived from scanning is representative only if every shot in the scan is ``typical'' (say within 1$\sigma$ of the expectation value) of an interaction pulse.  If the scan consists of 10 shots, the probability every shot is typical in this sense is 2.2\%.  At minimum, the statistics of relating such scanning measurements to the typical performance of the system must be analysed.  If the pulse profile is measured before full amplification, changes in the spectrum due to the amplification process can easily induce percent-level changes that are here shown to significantly affect outcomes.  That significance increases with the achieved peak intensity.  At $I\simeq 10^{23}$ W/cm$^2$, even a 1\% peak is already an order of magnitude more intense than the peak pulse in \cite{Malka:1997zwk}.  For precise experiments, an on-shot pulse temporal profile measurement at full amplification of every interaction pulse is therefore a requirement for future ultra-relativistic laser systems.

\subsection{Phase of injection}
Another reason Ref. \cite{McDonald:1998pu} may not apply is that the electron-laser interaction is finite duration, contrary to the setup of the authors' calculation.  A formally infinite interaction time in practice assumes the interaction begins early enough that the switch to laser-dominated dynamics does not occur suddenly.  There are two ways, common in laser-plasma experiments, this assumption can be violated.  First, the presence of nearby charges screens the laser field so that an individual charge is subject to a net force only after the laser field strength becomes greater than the typical electric field in the plasma $\langle \vec E^2\rangle_{pl}^{1/2}$, which varies greatly depending on the creation of the plasma.  This effect explains the success of plasma-mirror injection \cite{thevenet2016vacuum}, since electrons are released into free space (ejected from the plasma) near the peak laser intensity and thus see a sudden jump from nearly zero average force in plasma to large Lorentz force of the laser.
Second, in case the electrons arise from photoionization, the ionization threshold means that electrons become free only when the laser achieves a field strength comparable to the Coulomb field of the atom.  For hydrogen and helium plasmas, this effect is negligible, but for heavy inert gases such as argon or xenon, $^1S$ electrons may be liberated only within one or two decades of the peak intensity, which is the subject of considerable study, e.g. Refs. \cite{chowdhury2004electron,dichiara2008relativistic,yandow2019direct} and references therein.  Without having measured the initial distribution of electrons, the time (relative to the pulse's peak) the electron-laser interaction turns on must be considered uncontrolled.

A simple model of sudden turn-on is obtained by considering the net force on the particle zero until it passes a given threshold.  For simulations, we modify the profile function with a step-function $f(\varphi)\to f(\varphi)\Theta(\varphi+\varphi_c)$, with $\Theta(z>0)=1$ the Heaviside function.  With $\varphi_c>0$ the step occurs at $\varphi=-\varphi_c$. Modeling the transition to laser-dominated dynamics this way, preceding dynamics are incorporated into the (anyway unknown) initial distribution of electron positions and momenta.  If the threshold arises from screening, the electrons are free but have nonzero momentum, and we assume a maximum entropy distribution (Maxwellian).  

The sharp turn-on clearly violates \eq{profilecondition} as well as the slowly-varying condition inherent in the RPF \cite{Quesnel:1998zz,startsev1997multiple}, and we expect both a greater energy gain and a significant asymmetry in the transverse momentum distribution of the final particles.   As a consequence, the final particle energy and momentum distribution become sensitive to carrier-envelope phase and other non-ponderomotive effects.  The largest asymmetry arises when the electron is injected at a phase corresponding to a maximum of the electric field, and the final electron energy becomes less sensitive to the pulse profile function than the phase at which the interaction begins, in agreement with ionization studies \cite{yandow2019direct}. 

\begin{figure}[h]
\includegraphics[width=0.6\textwidth]{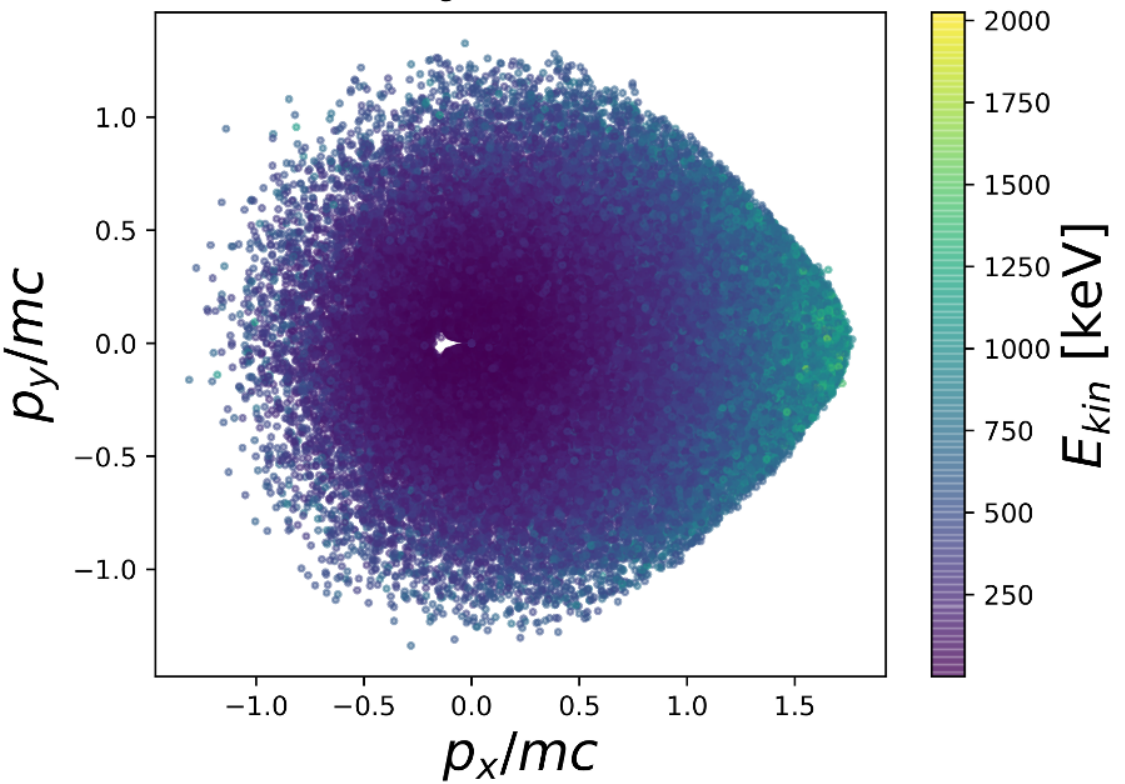}
\caption{\label{fig:pTphic} Scatter plot of final momenta for electrons scattered by a sech-profile pulse ($a_0=3$, $w_0=10\,\mu$m, $\Delta\tau=500$\,fs) and initial step at $\phi_c=0.7$.}
\end{figure}

\fig{pTphic} shows the distribution of final transverse momentum $|\vec p_T|=\sqrt{p_x^2+p_y^2}$ for $\varphi_c=0.7$, which is chosen among many possible values of $\varphi_c$ that can yield electrons with kinetic energy up to $\simeq 900$ keV.  Notice the $p_T$ distribution is not azimuthally symmetric with additional higher momenta in the $p_x>0$ direction, corresponding to particles that first enter the laser field near the focal spot where the laser fields are maximum.  As shown by the color-coding these particles also achieve the highest energy, consistent with our reasoning.  This effect is pronounced for photoionized electrons, which begin free propagation when the peak laser field is in focal region \cite{chowdhury2004electron}.  The anisotropy  arises from the maximum of the E field coinciding with the peak of the profile due to the specific value of the carrier-envelope phase in our model of the fields. 

Directly measuring when the free electron-laser interaction begins is difficult.  As ionization and plasma screening both originate in dynamics, determining the number of free electrons in the laser spot requires a probe with time resolution smaller than the high-intensity laser pulse.  For the foreseeable future, this parameter would have to be fit to the experimental data during data analysis.  Such fitting is addressed in part by ongoing efforts to observe strong-field ionization processes and validate theory calculations of the ionization rate.

\subsection{Initial transverse momentum}\label{sec:pT}
In the experiment of Ref. \cite{Malka:1997zwk}, the electrons were generated by laser-ablatation of a solid target situated off the axis of the acceleration laser's propagation.  The schematic in Fig. 1 of Ref. \cite{Malka:1997zwk} does not give the off-axis distance, but the fact that acceleration laser was focused with an f/3 OAP suggests that the angle from which the electrons entered the focal region was greater than tan$^{-1}(1/3)=18.4^\circ$.  Equivalently $p_\perp=\sqrt{p_x^2+p_y^2}\gtrsim p_z/3$.  

Although the RPF is azimuthally symmetric for a gaussian beam model of a laser field, the final distribution of electrons scattered by the laser is azimuthally symmetric if and only if the initial distribution is azimuthally symmetric.  Ref. \cite{Quesnel:1998zz} observed a ring in the $p_x$-$p_y$ plane (their Fig. 9) because they initialized the electrons in an infinitely thin disk centered on the beam axis with velocity colinear to the pulse propagation.  The initial transverse momentum present in the experiment breaks the azimuthal symmetry of the scattering, and a broader three-dimensional distribution of electron initial positions shows many electrons scattering into smaller $p_\perp$ final states.  Azimuthal symmetry will be visible for very large energy gain $\Delta E/E\gg 1$, because in this limit the initial state momentum is negligible compared to the final momentum.

To illustrate, we simulate electrons traveling toward the focal region at a $54^\circ$ angle from the beam axis.  The electrons are initially uniformly distributed off-axis between the laser pulse and focal plane and all have the same speed $|\vec v|=0.3c$.  The laser pulse has a gaussian temporal profile, peak intensity $I_0=$, focal spot size 20\,$\mu$m and duration 350 fs.  We verified that the same results are generated by solving either the RPF or the Lorentz force.

\begin{figure}[h]
\includegraphics[width=0.6\textwidth]{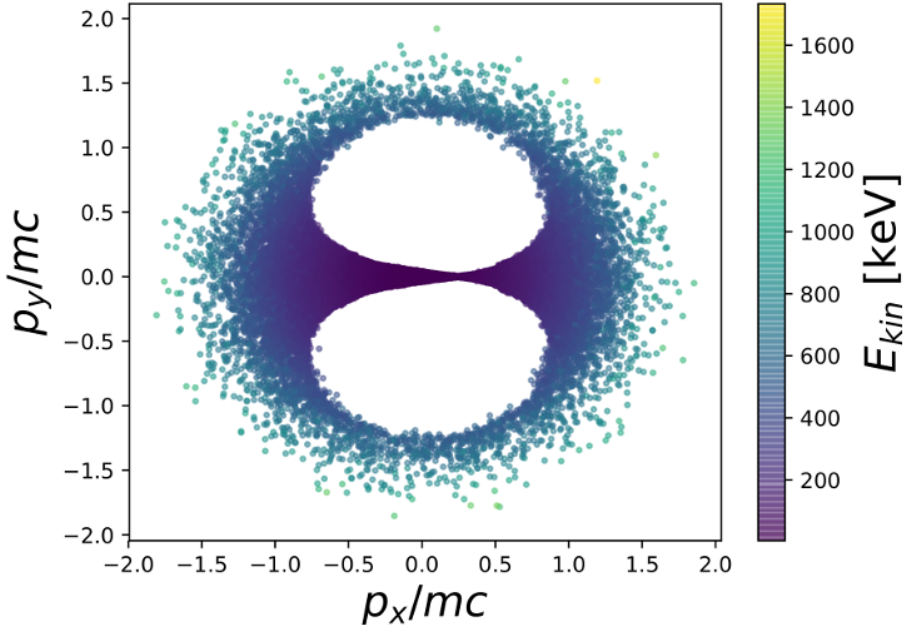}
\caption{\label{fig:pTpTi} Scatter plot of final momenta for electrons scattered by a sine-squared profile pulse ($a_0=3$, $w_0=20\,\mu$m, $\Delta\tau=350$\,fs).  Electrons have $|\vec v|=0.3c$ and enter the focal region at $54^\circ$ from the laser axis.}
\end{figure}

Most electrons remain in the injection plane, whether aligned with polarization or not.  Electrons that interact with only the edges of the pulse (in space or time) are accelerated less and hence less diverted from their initial momentum vectors. For this reason, it remains surprising that Ref. \cite{Malka:1997zwk} did not observe electrons upon rotating the plane of polarization; the electron source and detector remained in the same plane and only the laser's polarization changed by $90^\circ$.  This suggests that more than one revision to the model of the experiment will be necessary to fit the data.

The highest energy electrons are scattered more nearly isotropically, but only if the initial distribution of electrons accesses the full transverse distribution of the laser fields.  
If the electrons responsible for the detector signal arrive in the focal region before the pulse does, then they homogeneously sample the transverse intensity gradients, with different electrons seeing gradients in different radial directions.  On the other hand, if the electron energy and time-of-arrival at the focal region are correlated (e.g. due to dynamics at the source), then time-of-arrival becomes another model parameter.  

\begin{figure}[h]
\includegraphics[width=0.6\textwidth]{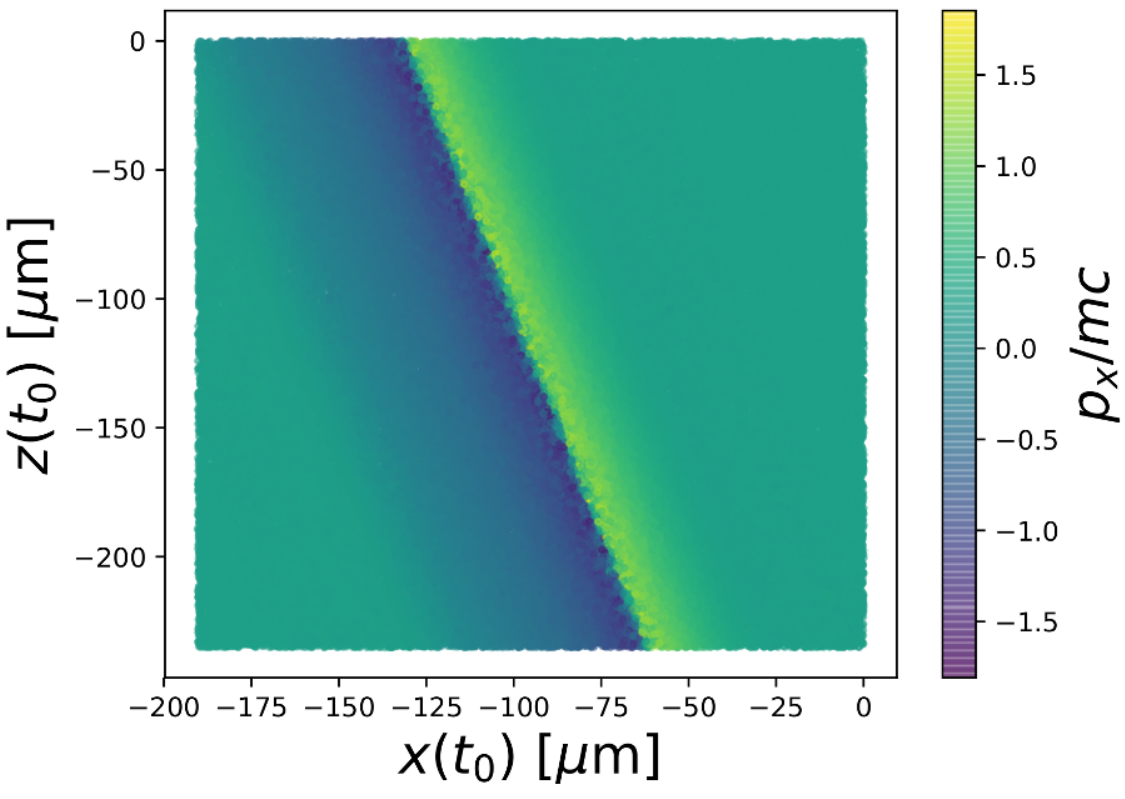}
\caption{\label{fig:diffx0}  In-plane momentum (normalized to mass, $p_x/m$) as a function of initial position for the simulation that above reproduced data of \cite{Malka:1997zwk} using initial transverse velocity.}
\end{figure}

\fig{diffx0} shows the final $p_x$ (in polarization plane) as a function of the initial position for the same parameters as \fig{pTpTi}.  With the magnitude of the velocity fixed for all particles, initial position completely determines the time-of-arrival.  Electrons arriving in the focal region ahead of the laser pulse scatter in the $+\hat x$ direction, because, with the highest intensity on axis and behind them, they experience a ponderomotive force in the $+\hat x$ direction.  In contrast, electrons arriving slightly later than the laser pulse, see the ponderomotive force pointing in the $-\hat x$ direction.  Therefore, a 100-fs variation in the source dynamics (much smaller than the ablation time scale in the experiment) enhances or suppresses the number of electrons reaching the detector, which was placed on only one side of the beam axis.

In future experiments, the source of electrons must be thoroughly characterized before data shots with the scattering laser on.  Moreover, this characterization must be accompanied by analysis of the uncertainties in order to distinguish the signal from random fluctuations of the source once the scattering laser is on.

\subsection{Shot-to-shot variations in the laser}\label{sec:shotvariation}

Laser systems fluctuate.  The energy input during amplification, the spectrum of the pulse and the pointing all vary from shot to shot.  These fluctuations mean the on-target pulse profile, focal spot size, and intensity distribution in space and time \cite{pariente2016space} differ from shot to shot.  Systematic changes could have a similar or greater impact: in Ref. \cite{Malka:1997zwk} changing the polarization of the laser with an optical element could result in uncontrolled changes to the wavefront, impacting both focus quality and pulse duration.  

The laser parameters given are not quite consistent with each other: a 20J, 300fs sine-squared temporal profile, spatially gaussian pulse with $w_0=10\,\mu$m has a peak intensity of $1.80\times 10^{19}$W/cm$^2$, corresponding to a dimensionless laser amplitude of $a_0\simeq 3.64$.  Increasing the pulse duration to 500fs implies a peak intensity of $1.08\times 10^{19}$W/cm$^2$ or $a_0\simeq 2.82$.  Refs. \cite{Malka:1997zwk} and \cite{Quesnel:1998zz} used $w_0=10\,\mu$m for simulations, but the experiment states that the pulse was focused into the target region by an f/3 optic, which for a 1 micron-wavelength laser suggests $w_0\simeq 2.62\,\mu$m \footnote{This is an approximation valid to about 5\%--the amount of energy in the higher order maxima of an Airy transverse profile as compared to a gaussian transverse profile with the same FHWM}.   For this $w_0$, the nominal peak intensities are $2.62\times 10^{20}$W/cm$^2$ for a 300-fs pulse and $1.58\times 10^{20}$W/cm$^2$ for a 500-fs.  The expected intensity depends on the choice of pulse profile function, differing by a factor of 2 between them.  The authors give a range for the pulse duration but not the energy or spot size, though all three certainly vary.

Many changes in the laser cannot be accounted for by the gaussian beam model used here.  Physically, most causes of shot-to-shot variances will result also in departures from the nominal model of the near-focus laser fields.  Since approximations such as the gaussian beam and RPF depend on the laser parameters, the accuracy of predictions also varies with the fluctuations in the laser.  Even so, varying the parameters of the gaussian beam model provides an estimate of the impact of shot-to-shot (statistical) variation in observables.

\begin{figure}[h]
\includegraphics[width=0.6\textwidth]{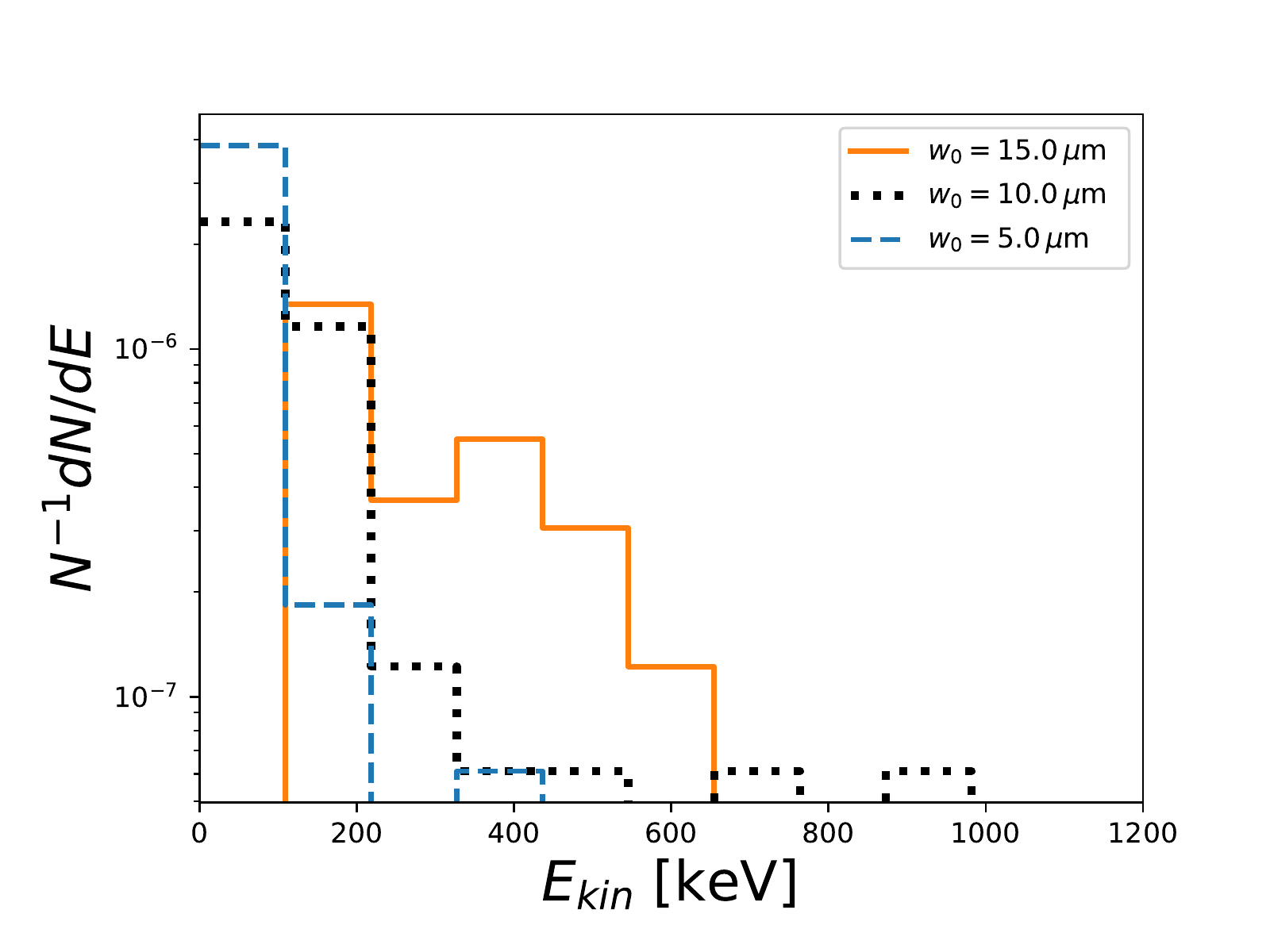}\caption{\label{fig:diffw}  Measured spectra along two sight lines, 39$^\circ$ and 46$^\circ$ from the beam axis for a 20J, 367fs laser pulse for varying waist size $w_0$. }
\end{figure}

We investigate the observable measured in Ref. \cite{Malka:1997zwk}: the spectrum of electrons along two sight lines, 39$^\circ$ and 46$^\circ$ from the beam axis.
Varying the pulse duration between 300 fs and 500fs changes the electron distributions relatively little, a factor 2-5 at higher energies and less at lower energies. Figure \ref{fig:diffw} shows the observable for the stated spot size (10$\mu$m) as well as smaller (5$\mu$m) and larger (5$\mu$m) spot sizes.  The laser energy is 20 J, pulse duration 367 fs and the temporal profile gaussian with the laser-electron interaction turning on at $\phi_c=1.2$.  The laser begins 20 Rayleigh lengths from the focal plane, and the electrons are initially uniformly distributed in a box between the laser and the focal plane extending 13 microns from the beam axis.  The electrons have initial momentum colinear to the laser propagation axis and initial energy following a Maxwellian distribution with temperature 10 keV.  Identical initial distributions of electrons are used in order to ensure that the relative probability of observing electrons in an given direction and energy bin is not affected by the density of particles in phase space.

An error or fluctuation in the spot size leads to an order of magnitude difference in the number of electrons detected.   Note in passing that the endpoint of the spectrum, the energy at which the electron distribution decreases below a fixed (e.g. detector) cutoff, does not correlate monotonically to spot size, being influenced by shot-to-shot variation.

Addressing the shot-to-shot variances of laser systems will require both new measurement techniques and data analysis techniques.  High intensity lasers are technically difficult to diagnose, often because the intensity and/or physical size of the beam is too great for well-established optical diagnostics.  In many cases the experimental setup precludes measuring laser parameters on-shot.  These facts require complementary development of statistical analysis of laser performance, especially issues such as the correlation between partial-beam measurements (e.g. of energy and temporal profile) to full-beam measurements.   

Some measurements, such as the complete spatio-temporal profile, are only possible in scanning mode with only a tiny fraction of the full pulse energy ($10^{-4}--10^{-6}$).  Single shot measurements do not exist, and therefore the variation between two single shots at full energy is unknown.  At higher peak intensities, even percent or per mille level fluctuations can have an effect.  Full energy shots at the largest laser systems  are essentially experiments in their own right.  Since experimental campaigns on large single-shot (100-1000J) class lasers consist of a few tens of shots at most, using a significant number of those for laser characterization would have a large impact on the actual experiment.  As laser performance parameters are liable to evolve over time, one would have to repeat this characterization for every experiment.  Only systematic development and constant calibration of new on-shot diagnostics can rectify this situation.

\section{Global fits}

We may now proceed to explain the data of Ref. \cite{Malka:1997zwk}.  The most difficult feature to fit is the absence of electrons perpendicular to the polarization plane.  To compare electron numbers in-plane of polarization to out-of-plane of polarization, the authors kept the apparatus fixed and rotated the polarization of laser.  However, this method actually changes two parameters, both the initial momentum vector and the observation line of sight  being rotated into the plane perpendicular to polarization.  Thus the experiment did not measure how much electrons are scattered out of the initial plane, which is defined by the initial momentum and laser propagation vectors.  As noted above (cf. Sec. \ref{sec:pT}), the RPF leads one to expect the highest energy electrons are scattered azimuthally symmetrically while lower energy electrons remain closer to the initial plane.  The absence of signal even at low energy suggests that the result may be partially due to different numbers of electrons from the source on different shots.

Nevertheless, we fit the data assuming the electron source remains consistent between shots to demonstrate the importance of the missing pieces of the model we identified above. Each of the four effects can enhance or suppress the number of electrons detected along the chosen sightline by a factor 10, and each alone is insufficient to explain the $\gtrsim$100-times difference in electron number between the in-plane and out-of-plane shots.  To obtain the $\sim$100-fold difference, we combine a change in laser spot size with either the sudden turn-on or the initial transverse momentum.  Although a simplistic model, varying $w_0$ may be the most relevant change in the laser since wavefront differences from rotating the polarization can easily translate into less energy enclosed in the FWHM central disk.

\begin{figure}[h]
\includegraphics[width=0.47\textwidth]{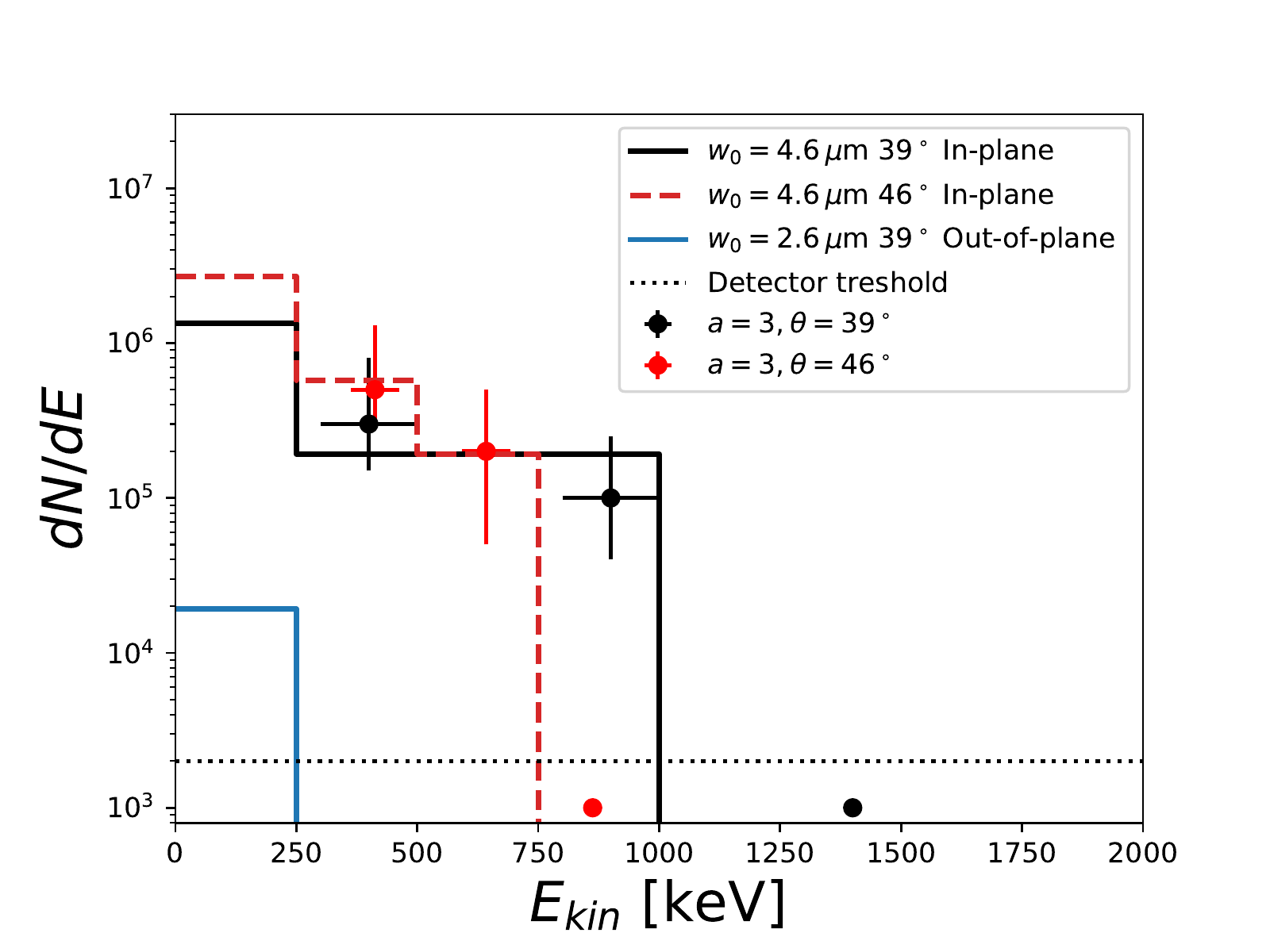}
\caption{\label{fig:globalbest1}  Measured spectra for a 20J, 367fs laser pulse, allowing $w_0$ to differ between in-plane ($w_0=4.6\,\mu$m) and out-of-plane ($w_0=2.6\,\mu$m) measurements. }
\end{figure}

The first fit is shown in Figure \ref{fig:globalbest1}.  In comparing to the data, we fix the laser energy, pulse duration and select an injection phase by fitting $dN/dE$ at $39^\circ$ and $46^\circ$ in the plane of polarization.  We set the overall normalization of the predicted $dN/dE$ to fit the data, since it reflects the total number of electrons in the experiment (which in turn depends on efficiency of the source) and is not part of our modeling.  We then vary the spot size to obtain a second prediction such that the out-of-polarization-plane $dN/dE$ prediction fits the (absence of) sigal.  We plot the predictions together with the data in the same manner as Fig. 2 of Ref. \cite{Malka:1997zwk}.  

For this first fit, we use a gaussian temporal profile. The specific parameters that fit the data depend on the pulse profile function, but we have found fitting parameters for each profile function.  

\begin{figure}[h]
\includegraphics[width=0.47\textwidth]{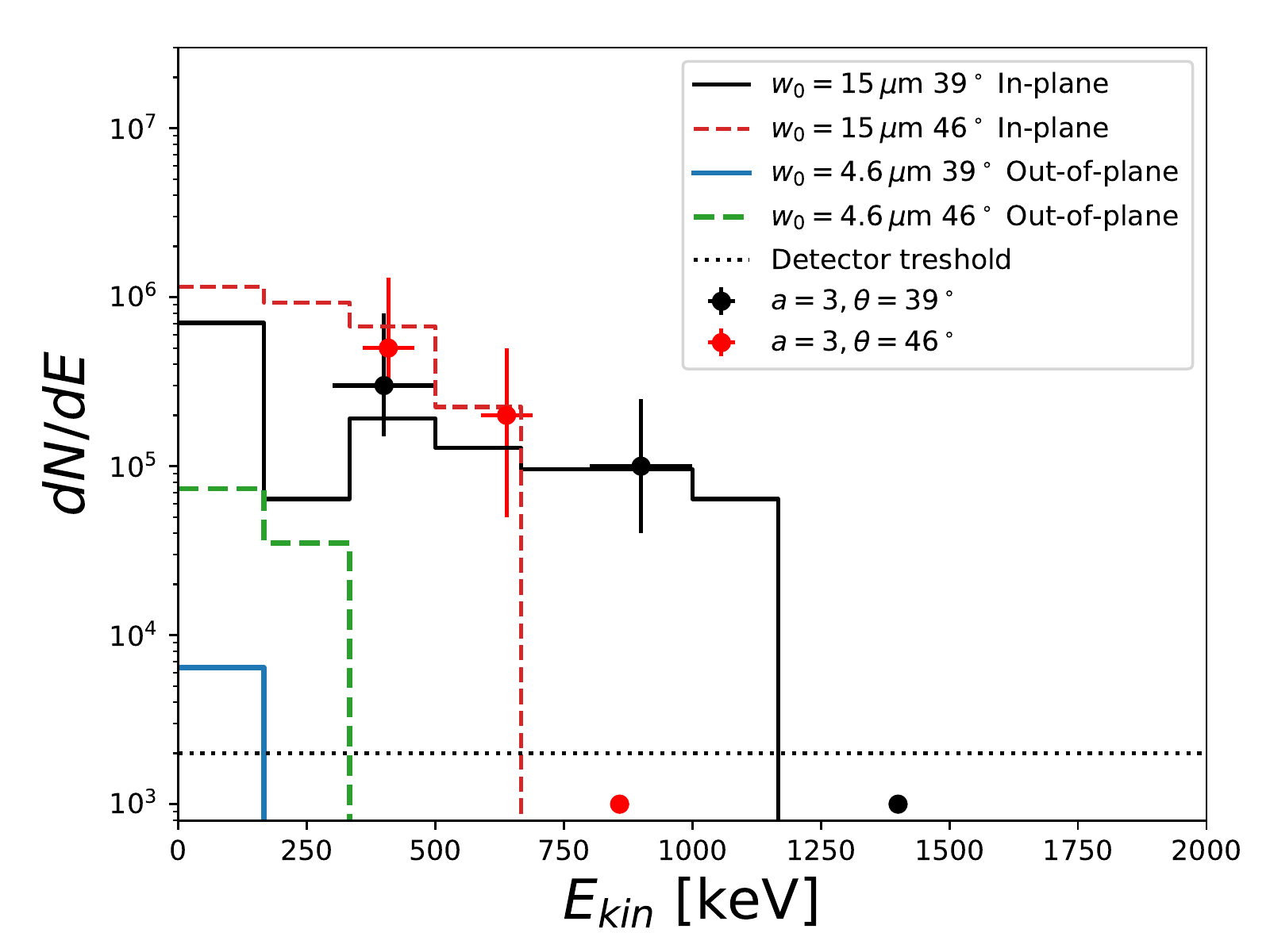}
\caption{\label{fig:globalbest2}  Measured spectra for a 16J, 350fs laser pulse, allowing $w_0$ to differ between in-plane ($w_0=15\,\mu$m) and out-of-plane ($w_0=4.6\,\mu$m) measurements. }
\end{figure}

In the second fit, \fig{globalbest2}, we follow the same procedure, choosing the laser parameters and initial transverse momentum to fit the in-plane measurements of $dN/dE$ and then varying the spot size to reproduce the relative suppression of out-of-plane electron number. In this case, we use a sine-squared temporal profile, which yields higher energy electrons without any sudden turn-on (recall \fig{profiles}).

Quantifying the goodness of fit is not productive in this case, because even this almost-minimal modeling has more parameters than the data has points to fit.  Qualitative comments are in order though: the goodness of the fit depends also on how the simulation and experiment data are binned.  We have chosen bin sizes roughly equal to the displayed uncertainty in the energy measurements presented in Ref. \cite{Malka:1997zwk}.  If we reduce the bin size, more information in the spectrum would become visible, which might provide more information about the physics of the interaction.  However, even constrained to the bin size suggested by the experimental resolution, the fit can be improved or degraded by changing bin size.

\section{Conclusions}

We have shown that more careful analysis and modeling can resolve a long-standing disagreement as to whether laser acceleration of electrons in free space was observed.  Differences between application of the theory and the experiment sufficed to change qualitatively the predicted electron distributions and thus obscured the physics in the experiment.  We found multiple examples of parameters that yielded predictions consistent the experimental data, but too little information makes it impossible to fit the data uniquely.  We cannot conclude with an overall ``best fit'' interpretation of the data.  

Our results help unify understanding of the various experiments seeking laser acceleration in free space.  Very long laser pulses interacting with low-energy free electrons result in very little energy gain \cite{McDonald:1998pu} but do increase electrons' energy spread \cite{cline2013first,carbajo2015direct}.  Sudden turn-on of the laser-electron interaction, whether due to ionization near the peak intensity \cite{mcnaught1998photoelectron, moore1999laser, dichiara2008relativistic,payeur2012generation, yandow2019direct} or ejection from a dense plasma \cite{thevenet2016vacuum}, can yield much higher energy gain.  Sudden turn-on makes the electron energy less sensitive to the temporal profile and more sensitive to precise magnitude of the field and initial momentum distribution of the electrons at the beginning of the interaction.

This exercise also suggests several lessons for future experiments.  Experiments must provide a quantitative description of the initial plasma or electron distribution, and measure the laser's shot-to-shot variations in performance.  Knowing the laser pulse's temporal profile is important when the electrons are likely to see the early-time features, whether a slow exponential turn-on (like a sech profile) or an extra peak in the pre-pulse.  It is less important when the electrons start to interact with the laser later in the pulse, particularly near the peak.  With such measurements in hand, we can try to fit the more-difficult-to-measure parameters, such as the time at which the free particle-laser interaction begins.  Since some laser parameters cannot be measured on-shot (such as laser spot size), we also need new methods to utilize off-shot measuremnts of spot size in modeling and data analysis.

Developing these measurement and data analysis procedures is essential moving forward.  Accurate single-electron dynamics are crucial to validating theories of radiation reaction and strong-field ionization, particularly as contemporary experiments continue to much higher intensity.  It is worth demonstrating the (increasing) necessity of accurate knowledge of laser parameters as laser intensity increases.  In order to quantify the typical error from incorrect or fluctuating laser parameters, we introduce a crude error functional,
\begin{align}\label{eq:errormeasure}
\Delta &= \bar{N}^{-1}\int \left|\frac{dN^{(1)}}{dE}-\frac{dN^{(2)}}{dE}\right|dE \notag\\
\bar{N} &= \frac{1}{2}\int \left(\frac{dN^{(1)}}{dE}+\frac{dN^{(2)}}{dE}\right)dE=\frac{1}{2}(N^{(1)}+N^{(2)}),
\end{align}
which compares the predictions of $dN^{(i)}/dE$ for sets (1) and (2) of laser and plasma initial conditions and outputs a number.  For example, for the two measurements of $dN/dE$ at $39^\circ$ and $46^\circ$ in \fig{globalbest2}, $\Delta$ is 1.98 and 1.87, respectively, showing that an error in this range corresponds to a nearly 0 or 1 difference in signal. \eq{errormeasure} can be experimentally measured, provided a laser system with sufficiently precise control to run statistically distinct experiments at the parameter sets (1) and (2).  However, it is neither a standard measure of the error nor a prediction interval.  We will discuss rigorous statistical methods in a later work.  For this work, \eq{errormeasure} provides a simple measure of the difference between a predicted electron spectrum and a measured result.

\begin{table}
\begin{tabular}{l|c c c c c}
Facility & Energy [J] & Duration [fs] & $w_0$ [$\mu$m]  & $I_0$ [W/cm$^2$] & $a_0$ (peak) \\\hline
Malka 1 \cite{Malka:1997zwk} & 20 & 300 & 2.6 & $4.25\times 10^20$ & 17.7 \\\hline
Malka 2 & 20 & 50 & 2.6 & $2.55\times 10^{21}$ & 43.3 \\\hline
Berkeley cite{} & 40 & 30 & 1.8 & $1.77\times 10^{22}$ & $91.4$ \\\hline
TPW f/3 low-P & 50 & 150 & 2.6 & $2.13\times 10^{21}$ & 39.55 \\\hline
TPW f/1 & 150 & 150 & 1.25 & $2.76\times 10^{22}$ & 142.5 \\\hline
CoReLS & 50 & 30 & 1.8 & $2.22\times 10^{22}$ & 127.7 \\\hline
OPAL & 600 & 20 & 1.25 & $8.28\times 10^{23}$ & 702.4
\end{tabular}
\caption{The pulse duration is measured as intensity full-width half-max.  The pulse profile is always gaussian.  Malka 1 uses the laser energy and pulse duration of Ref. \cite{Malka:1997zwk} with the nominal best focus for an f/3 mirror.  Malka 2 has mostly the same laser parameters as Malka 1 but is compressed to 50 fs, as would be possible with current laser technology. \label{tab:facilities}}
\end{table}

We compute $\Delta$ for nominal parameters for current and planned major high-intensity laser facilities, given in Table \ref{tab:facilities}.  We compare predictions for two different values of the laser spot size $w_0$, one the nominal focal spot size given in the table and second a focal spot twice as large.  Other laser parameters are the same between parameter sets (1) and (2).  

\begin{figure}[h]
\includegraphics[width=0.6\textwidth]{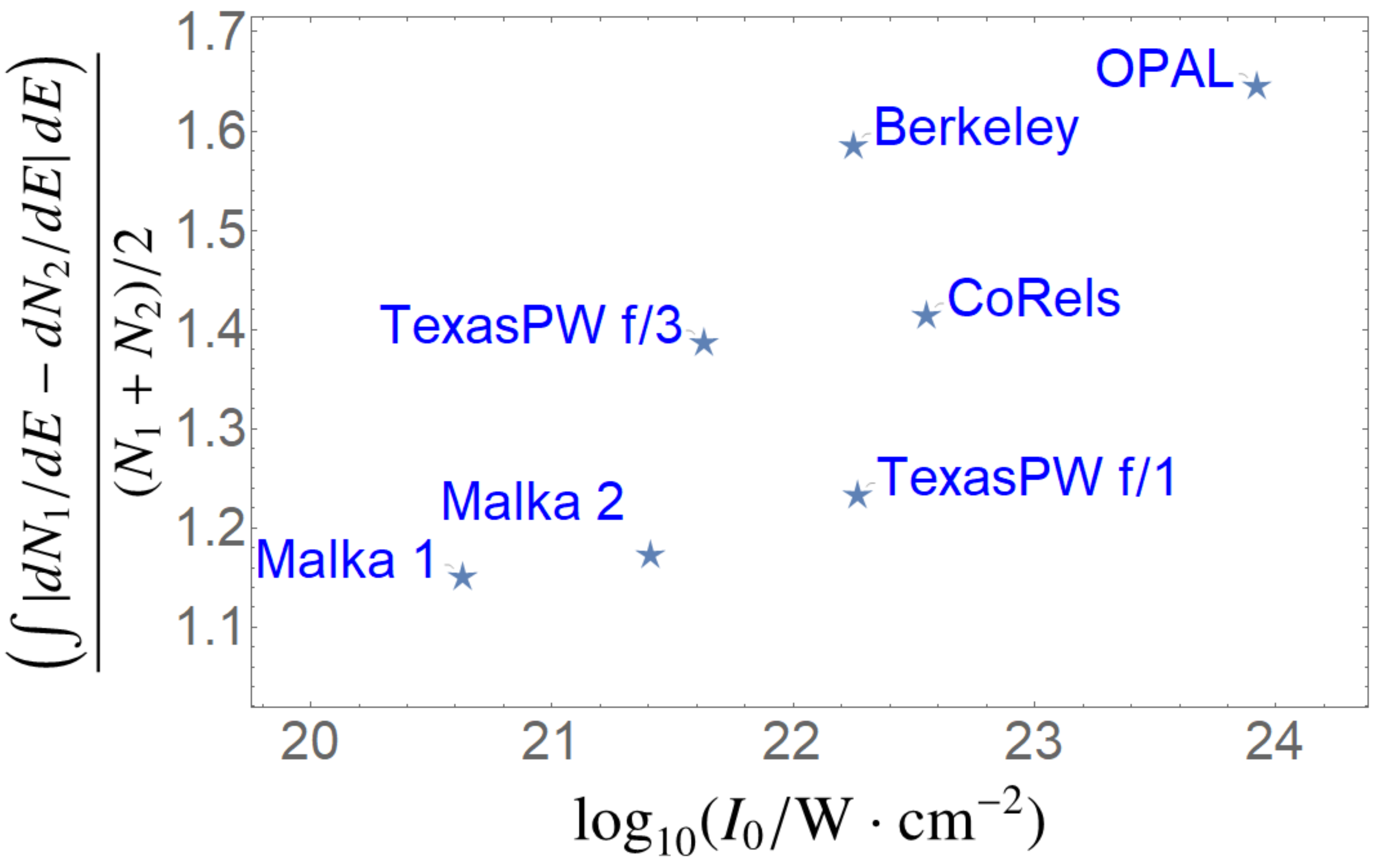}
\caption{\label{fig:w0errorvI0}  The error measure $\Delta$ for current and future laser facilities' nominal operating parameters in Table \ref{tab:facilities}.  For each point plotted at the facility's nominal peak intensity, laser energy is constant while the waist size $w_0$ increased by a factor 2.  The baseline value of $\Delta$ for macroscopically identical simulations is 0.2.}
\end{figure}

The difference in the observables for different laser parameters is large; the baseline value of $\Delta$ for two simulations with the same laser parameters but different (randomized) electron positions and momenta is less than 0.2.  $\Delta$ increases with intensity, suggesting that uncertainty arising unknown laser parameters becomes a more serious problem on higher-intensity laser systems. 

Since desirable physics goals, such as radiation reaction, are perturbations to the classical dynamics, discovery is impossible without a quantitative understanding of the baseline.  Without improved experimental and data-analytic efforts, experiments are in danger of confusing a shot-to-shot fluctuation in laser performance with signal.

\bibliography{plasma}

\end{document}